\documentclass[aps,showpacs,superscriptaddress,nofootinbib,preprint]{revtex4}
\usepackage{graphicx}% Include figure files
\def\slashchar#1{\setbox0=\hbox{$#1$}
   \dimen0=\wd0 \setbox1=\hbox{/} \dimen1=\wd1
   \ifdim\dimen0>\dimen1 \rlap{\hbox to \dimen0{\hfil/\hfil}} #1
   \else  \rlap{\hbox to \dimen1{\hfil$#1$\hfil}} / \fi}
\def\p{\slashchar{p}}
\def\q{\slashchar{q}}
\def\D{\slashchar{D}}

\begin{document}
\title{ Weak Kaon Production off the Nucleon}

\author{M. Rafi \surname{Alam}}
\affiliation{Department of Physics, Aligarh Muslim University, Aligarh-202 002, India}
\author{I. \surname{Ruiz Simo}}
\affiliation{Departamento de F\'\i sica Te\'orica and IFIC, Centro Mixto
Universidad de Valencia-CSIC, Institutos de Investigaci\'on de
Paterna, E-46071 Valencia, Spain}
\author{M. Sajjad \surname{Athar}}
\affiliation{Department of Physics, Aligarh Muslim University, Aligarh-202 002, India}
\author{M. J.  \surname{Vicente Vacas}}
\affiliation{Departamento de F\'\i sica Te\'orica and IFIC, Centro Mixto
Universidad de Valencia-CSIC, Institutos de Investigaci\'on de
Paterna, E-46071 Valencia, Spain}

\begin{abstract}
The  weak kaon production off the nucleon induced by neutrinos is studied  at the low and intermediate energies
of interest for some ongoing and future neutrino oscillation experiments. This process is also potentially important for the analysis of proton decay experiments. We develop a microscopical model based on the SU(3) chiral Lagrangians. The basic parameters of the model are $f_\pi$, the pion decay constant, Cabibbo's angle, the proton and neutron magnetic moments and the  axial vector coupling constants for the baryons octet, $D$ and $F$, that are obtained from the analysis of the semileptonic decays of neutron and hyperons. The studied mechanisms are the main source of kaon production for neutrino energies up to 1.2 to 1.5 GeV for the various channels and the cross sections are large enough to be amenable to be measured by experiments such as Minerva and T2K.

\end{abstract}
\maketitle
\section{Introduction}
With the recent developments of the atmospheric and accelerator based neutrino experiments it is now well known that neutrinos oscillate and have finite masses.  Now, the main goal is to precisely determine  the different parameters of the Pontecorvo-Maki-Nakagawa-Sakata (PMNS) matrix, absolute neutrino masses, CP violating phase $\delta$, etc. The neutrino energy region of a few GeV  is quite sensitive to  the neutrino oscillation parameters. Therefore, most of the present experiments like MiniBooNE, K2K, T2K, No$ \nu $A, etc. have taken data or have been planned in this energy region. Neutrino detection proceeds basically through various channels of interaction with hadronic targets like quasielastic scattering, meson production, resonance excitations, etc...
Therefore, a reliable estimate of these cross sections has become  important.
There is a considerable ongoing  theoretical and experimental effort addressing this question (see e.g. the proceedings of the NUINT conference series~\cite{nuint}) with many of the studies concentrated at  low energies where
quasielastic scattering and  pion production dominate or in the deep inelastic scattering.
However, in the discussed  energy region, other not so well known processes like kaon and hyperon production may also become important. In principle, their cross sections are smaller than for the pionic processes because of phase space and the Cabibbo suppression for $\Delta S=1$ reactions. Nonetheless, in the coming years of precision neutrino physics, their knowledge could be relevant for the data analysis, apart from their own intrinsic interest related to the role played by the strange quarks in  hadronic physics.

The currently available data is restricted to a few events measured in bubble chamber experiments~\cite{Barish:1978pj,Baker:1981tx,Barish:1974ye}. However, this is expected to change soon. In particular,
MINER$\nu$A, a dedicated experiment to measure neutrino nucleus cross section using several nuclear targets like Carbon, Iron and Lead in the neutrino energy region of 1-20 GeV has recently started taking data. It is also planned to study specifically the strange particle production and it is expected that thousands of events would be accumulated where a kaon is produced in the final state~\cite{Solomey:2005rs}.

On the theoretical side there are  very few calculations which deal with  strange particle production at low neutrino energies: single hyperon production~\cite{VicenteSingh,Mintz:2007zz},
the study of several kaon and hyperon production channels of Dewan~\cite{Dewan}
and the work of Shrock~\cite{Shrock} who has analysed the $\Delta S=0$ processes.
At higher energies, Amer has studied the strange particle production assuming the dominance of s-channel resonant mechanisms~\cite{Amer:1977fy}.  
Also, in part as a consequence of the scarcity of theoretical work, the MonteCarlo generators used in the analysis of the experiments apply models that are not too well suited to describe the strangeness production at low energies. For instance, NEUT, used by Super-Kamiokande, K2K, SciBooNE and T2K, only considers associated kaon production  implemented by  a model based on the excitation and later decay of resonances~\cite{Hayato:2009zz}. A similar model is used by other event generators like NEUGEN~\cite{Gallagher:2002sf}, NUANCE~\cite{Casper:2002sd} (see also discussion in Ref.~\cite{Zeller:2003ey}) and GENIE~\cite{Andreopoulos:2009rq}. As it will be emphasized below, this approach is not appropriate for low energies strangeness production.

In neutrino induced reactions, the first inelastic reaction creating strange quarks is the single kaon production
(without accompanying hyperons)\footnote{For antineutrinos the lowest threshold for $|\Delta S|=1$ reactions is much lower and corresponds to hyperon production.}.
This  charged current (CC) $\Delta S=1$ process is particularly appealing for several reasons. One of them is the important background that could produce, due to atmospheric neutrino interactions, in the analysis of 
one of the main decay channels the proton has in many SUSY GUT models ($ p \rightarrow \nu + K^+ $)
\cite{Mann:1986ht,MarrodanUndagoitia:2006qn,Kobayashi:2005pe}. A second reason is its simplicity from a theoretical point of view. At low energies, it is possible to obtain model independent predictions using Chiral Perturbation Theory ($\chi$PT) and due to the absence of $S=1$ baryonic resonances, the range of validity of the calculation could be extended to higher energies than for other channels. Furthermore, the kaon associated production (with  accompanying hyperons)  has a higher energy threshold (1.10 vs. 0.79 GeV). This implies that even when the associated production  is not Cabibbo suppressed, for a wide energy region (such as the ANL, the MiniBooNE or the T2K neutrino spectrum) single kaon production could still be dominant~\cite{Dewan}.

The paper is organized as follows. In Sec.~\ref{sec:form} we present the formalism for CC single kaon production in  neutrino nucleon scattering based on the Lagrangians of SU(3) $\chi$PT. We also discuss the differences with previous calculations. Results, discussions  and our concluding remarks are presented in Sec.~\ref{sec:res}.

\section{Formalism}
\label{sec:form}
The basic reaction for the neutrino induced charged current kaon production  is
\begin{equation}\label{reaction}
\nu_{l}(k) + N(p) \rightarrow l(k^{\prime}) + N^\prime(p^{\prime}) + K(p_{k}),
\end{equation}
where $l=e,\mu$ and $ N \& N^\prime $=n,p.	
The expression for the differential cross section in the laboratory (lab) frame for the above process is given by,
\begin{eqnarray}\label{d9_sigma}
d^{9}\sigma &=& \frac{1}{4 M E(2\pi)^{5}} \frac{d{\vec k}^{\prime}}{ (2 E_{l})} \frac{d{\vec p\,}^{\prime}}{ (2 E^{\prime}_{p})} \frac{d{\vec p}_{k}}{ (2 E_{K})} \delta^{4}(k+p-k^{\prime}-p^{\prime}-p_{k})\bar\Sigma\Sigma | \mathcal M |^2,\nonumber
\end{eqnarray}
where  $ \vec{k}$ and $ \vec{k^\prime} $ are the 3-momenta of the incoming and outgoing leptons in the lab frame with energy $E$ and $ E^\prime$ respectively. The kaon lab momentum is $\vec{p}_k $ having energy $ E_K  $, $M$ is the nucleon mass,
$ \bar\Sigma\Sigma | \mathcal M |^2  $ is the square of the transition amplitude matrix element averaged(summed) over the spins of the initial(final) state. At low energies, this amplitude can be written in the usual form as
\begin{equation}
\label{eq:Gg}
 \mathcal M = \frac{G_F}{\sqrt{2}}\, j_\mu^{(L)} J^{\mu\,{(H)}}=\frac{g}{2\sqrt{2}}j_\mu^{(L)} \frac{1}{M_W^2}
\frac{g}{2\sqrt{2}}J^{\mu\,{(H)}},
\end{equation}
 where $j_\mu^{(L)}$ and $  J^{\mu\,(H)}$ are the leptonic and hadronic currents respectively, 
$G_F=\sqrt{2} \frac{g^2}{8 M^2_W}=1.16639(1)\times 10^{-5}\,\mbox{GeV}^{-2}$ is the Fermi constant and 
$g$ is the gauge coupling.
The leptonic current can be readily obtained from the standard model  Lagrangian coupling the $W$ bosons to the leptons 
\begin{equation}
{\cal L}=-\frac{g}{2\sqrt{2}}\left[{ W}^+_\mu\bar{\nu}_l
\gamma^\mu(1-\gamma_5)l+{ W}^-_\mu\bar{l}\gamma^\mu
(1-\gamma_5)\nu_l\right]=-\frac{g}{2\sqrt{2}}\left[j^\mu_{(L)}{ W}^+_\mu+h.c.\right].
\end{equation}

\begin{figure}
\begin{center}
\includegraphics[height=6cm, width=12cm]{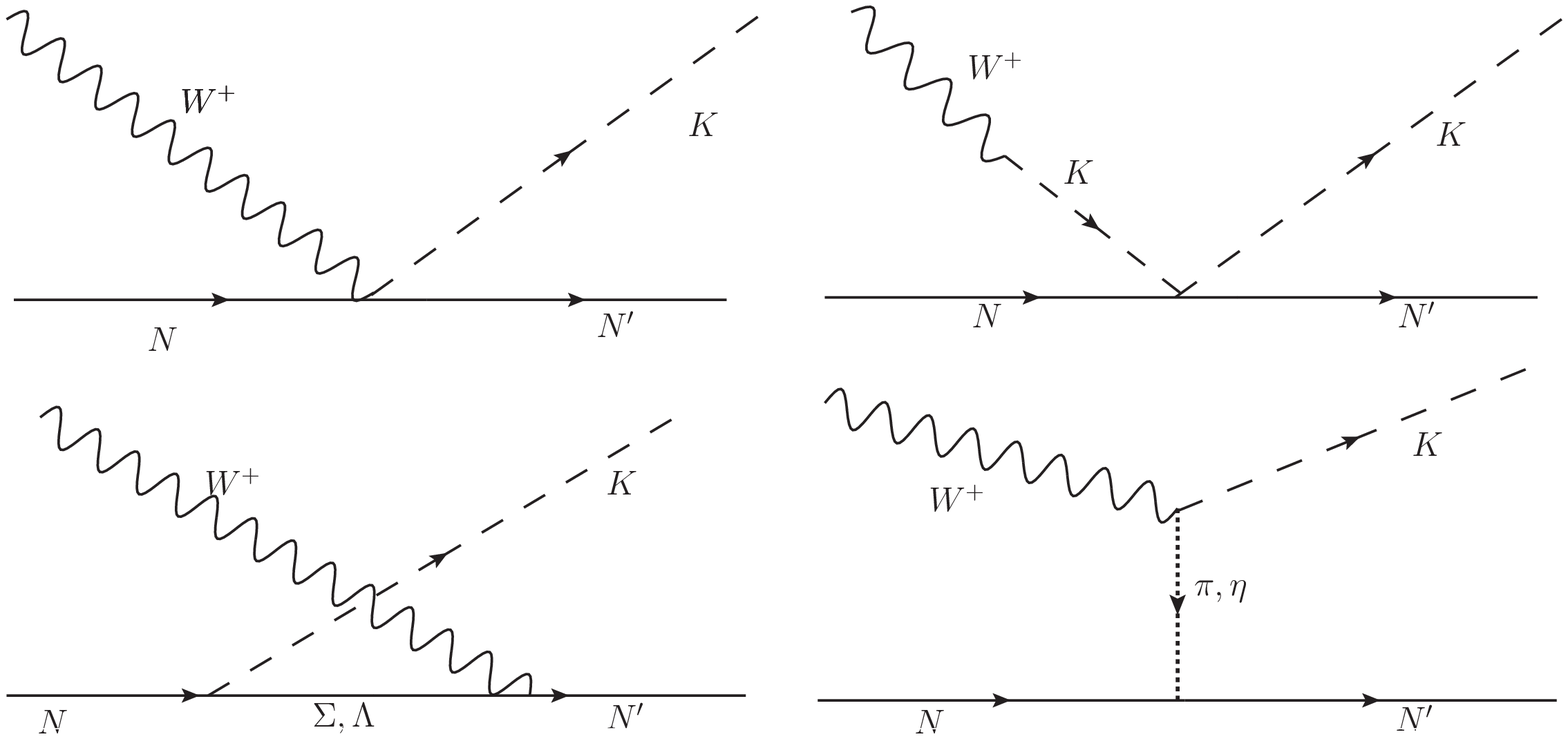}
\caption{Feynman diagrams for the process $ \nu N\rightarrow l N^\prime K $.  First row from left to right:  contact term (labeled CT in the text), Kaon pole term (KP); second row: u-channel diagram
($ Cr\Sigma$, $Cr\Lambda $) and Pion(Eta) in flight ($ \pi P$,  $ (\eta P) $ }
\label{fg:terms}
\end{center}
\end{figure}

We consider four different channels that  contribute to the hadronic current. They are depicted in Fig.~\ref{fg:terms}. There is a contact term (CT), a kaon pole (KP) term, a u-channel process with a $\Sigma$
or $\Lambda$ hyperon in the intermediate state and finally a meson ($\pi,\,\eta$) exchange term. For the specific reactions under consideration, there are not s-channel contributions given the absence of $S=1$ baryonic resonances. The current of the KP term is  proportional to $q^\mu$. This implies, after contraction with the leptonic tensor, that the amplitude is proportional to the lepton mass and therefore very small. 

%*****************************************************************
%*****************************************************************
The contribution of the different terms can be obtained in a systematic manner using $\chi$PT. This allows to identify some terms that were missing in the approach of Ref.~\cite{Dewan} which only included the u-channel diagrams in the calculation. The lowest-order SU(3) chiral Lagrangian describing the pseudoscalar mesons in the presence of an external current is
\begin{equation}
\label{eq:lagM}
{\cal L}_M^{(2)}=\frac{f_\pi^2}{4}\mbox{Tr}[D_\mu U (D^\mu U)^\dagger]
+\frac{f_\pi^2}{4}\mbox{Tr}(\chi U^\dagger + U\chi^\dagger),
\end{equation}
where the parameter $f_\pi=92.4$MeV is the pion  decay constant, $U$ is the SU(3) representation of the meson fields 
\begin{eqnarray}
U(x)&=&\exp\left(i\frac{\phi(x)}{f_\pi}\right),\nonumber\\
\phi(x)&=&
\left(\begin{array}{ccc}
\pi^0+\frac{1}{\sqrt{3}}\eta &\sqrt{2}\pi^+&\sqrt{2}K^+\\
\sqrt{2}\pi^-&-\pi^0+\frac{1}{\sqrt{3}}\eta&\sqrt{2}K^0\\
\sqrt{2}K^- &\sqrt{2}\bar{K}^0&-\frac{2}{\sqrt{3}}\eta
\end{array}\right),
\end{eqnarray}
and $D_\mu U$ is its covariant derivative
\begin{eqnarray}
D_\mu U&\equiv&\partial_\mu U -i r_\mu U+iU l_\mu\,.
\end{eqnarray}
Here, $l_\mu$ and $r_\mu$ correspond to left and right handed currents, that for the CC case are given by
\begin{equation}
r_\mu=0,\quad l_\mu=-\frac{g}{\sqrt{2}}
({W}^+_\mu T_+ + {W}^-_\mu T_-),
\end{equation}
with $W^\pm$ the $W$ boson fields and
$$
T_+=\left(\begin{array}{rrr}0&V_{ud}&V_{us}\\0&0&0\\0&0&0\end{array}\right);\quad
T_-=\left(\begin{array}{rrr}0&0&0\\V_{ud}&0&0\\V_{us}&0&0\end{array}\right).
$$
Here,  $V_{ij}$ are the elements of the 
Cabibbo-Kobayashi-Maskawa  matrix. 
The second term of the Lagrangian of Eq.~\ref{eq:lagM}, that incorporates the
explicit breaking of chiral symmetry coming from the quark masses~\cite{Scherer:2002tk}, is not relevant for our study.

The lowest-order  chiral Lagrangian for the baryon octet in the presence of an external current can be written in terms of the SU(3) matrix
\begin{equation}
B=
\left(\begin{array}{ccc}
\frac{1}{\sqrt{2}}\Sigma^0+\frac{1}{\sqrt{6}}\Lambda&\Sigma^+&p\\
\Sigma^-&-\frac{1}{\sqrt{2}}\Sigma^0+\frac{1}{\sqrt{6}}\Lambda&n\\
\Xi^-&\Xi^0&-\frac{2}{\sqrt{6}}\Lambda
\end{array}\right)
\end{equation}
as 
\begin{equation}
\label{eq:lagB}
{\cal L}^{(1)}_{MB}=\mbox{Tr}\left[\bar{B}\left(i\D
-M\right)B\right]
-\frac{D}{2}\mbox{Tr}\left(\bar{B}\gamma^\mu\gamma_5\{u_\mu,B\}\right)
-\frac{F}{2}\mbox{Tr}\left(\bar{B}\gamma^\mu\gamma_5[u_\mu,B]\right),
\end{equation}
where $M$ denotes the mass of the baryon octet, and the parameters $D=0.804$ and $F=0.463$
can be determined from the baryon semileptonic decays~\cite{Cabibbo:2003cu}.
The covariant derivative of $B$ is given by
\begin{equation}
D_\mu B=\partial_\mu B +[\Gamma_\mu,B],
\end{equation}
with
\begin{equation}
\Gamma_\mu=\frac{1}{2}\left[u^\dagger(\partial_\mu-ir_\mu)u
+u(\partial_\mu-il_\mu)u^\dagger\right],
\end{equation}
where  we have introduced $u^2=U$. Finally, 
\begin{equation}
u_\mu= i\left[u^\dagger(\partial_\mu-i r_\mu)u-u(\partial_\mu-i
l_\mu)u^\dagger\right].
\end{equation}
The next order meson baryon Lagrangian contains many new terms (see for instance Ref.~\cite{Oller:2006yh}). Their importance for kaon production will be small at low energies and there are some uncertainties in the coupling constants. Nonetheless, for consistency with previous calculations, we will include the contribution to the weak magnetism coming from the pieces 
\begin{equation}
{\cal L}^{(2)}_{MB}= d_5 \mbox{Tr}\left(\bar{B}[f_{\mu\nu}^+,\sigma^{\mu\nu}B]\right)+
d_4 \mbox{Tr}\left(\bar{B}\{f_{\mu\nu}^+,\sigma^{\mu\nu}B\}\right)+\dots,
\end{equation}
where the tensor $f_{\mu\nu}^+$ can be reduced for our study to
\begin{equation}
 f_{\mu\nu}^+=\partial_\mu l_\nu-\partial_\nu l_\mu -i [l_\mu,l_\nu].
\end{equation}
In this case, the coupling constants are fully determined by the proton and neutron anomalous magnetic moments.
This same approximation has also been used in calculations of single pion production induced 
by neutrinos~\cite{Nieves}. 
Now, writing the amplitude for the coupling of the $W$ boson to the hadrons for each of the terms in the form 
$\frac{g}{2\sqrt{2}} (J^\mu_{H}{ W}^+_\mu+h.c.)$, for consistency with Eq.~\ref{eq:Gg}, we get the following contributions to the hadronic current
\begin{eqnarray}
j^\mu \big|_{CT} &=& -i A_{CT} V_{us} \frac{\sqrt{2}}{2 f_\pi} \bar N(p^\prime) ( \gamma^\mu+
\gamma^\mu \gamma^5 B_{CT} ) N(p),\nonumber\\ 
j^{\mu}\big|_{Cr\Sigma} &=& i A_{Cr\Sigma} V_{us} \frac{\sqrt{2}}{2 f_\pi} \bar N(p^\prime)
 \left( \gamma^\mu 
+i\frac{\mu_p+2\mu_n}{2M}\sigma^{\mu\nu}q_\nu
+ (D-F)(\gamma^\mu-\frac{q^\mu}{q^2-M_k^2}\q )\gamma^5 \right)\nonumber\\
& &\times  \frac{\p - \p_k + M_\Sigma}{( p -  p_k)^2 -M_\Sigma^2} \p_k \gamma^5  N(p)  ,\nonumber\\
j^{\mu}\big|_{Cr\Lambda}&=&i A_{Cr\Lambda} V_{us} \frac{\sqrt{2}}{4 f_\pi} \bar N(p^\prime)\left( \gamma^\mu
 +i\frac{\mu_p}{2M}\sigma^{\mu\nu}q_\nu
-\frac{D+3F}{3} (\gamma^\mu -\frac{q^\mu}{q^2-M_k^2}\q )\gamma^5  \right)\nonumber\\
& &\times \frac{\p - \p_k +M_\Lambda}{( p -  p_k)^2 -M_\Lambda^2}   \p_k \gamma^5 N(p) ,\nonumber \\
j^{\mu}\big|_{KP}&=& i A_{KP} V_{us} \frac{\sqrt{2}}{4 f_\pi} \bar N(p^\prime) (\q +\p_k) N(p)  \frac{1}{q^2-M_k^2}
  q^\mu,\nonumber \\
j^{\mu}\big|_{\pi}&=& i A_{\pi P} V_{us} (D+F) \frac{\sqrt{2}}{2 f_\pi} \frac{M}{(q-p_k)^2 - {M_{\pi}^2}} \bar N(p^\prime) \gamma^5.(q^\mu - 2 {p_k}^\mu) N(p),\nonumber \\
j^{\mu}\big|_{\eta }&=& i A_{\eta P} V_{us} (D-3F) \frac{\sqrt{2}}{2 f_\pi} \frac{M}{(q-p_k)^2 - {M_{\eta}^2}} \bar N(p^\prime)\gamma^5.(q^\mu - 2 {p_k}^\mu) N(p), 
\end{eqnarray}
where, $q=k-k^\prime$ is the four momentum transfer, $ V_{us}=\sin{\theta}=0.22 $ where $ \theta $ is the Cabibbo angle,
$N(\cdotp),\, \bar N(\cdotp)$ denote the nucleon spinors, $\mu_p=1.7928$ and  $\mu_n=-1.9130$ are the proton and neutron anomalous magnetic moments. The value of the various parameters of the formulas are shown in Table~\ref{tab:1}. 
\begin{table*}
\begin{center}
\caption{Values of the parameters appearing in the hadronic currents.}
\label{tab:1}
%\resizebox{148mm}{40mm}{
\begin{tabular}{|c|c c c c c c c| }\hline\hline
Process & $A_{CT} $&$ B_{CT}$ & $A_{Cr\Sigma}$ & $A_{Cr\Lambda}$ & $A_{KP}$ &$ A_{\pi P}$ & $A_{\eta P}$ \\ \hline\hline
$ \nu n\rightarrow lKn$ & 1 &D-F& -(D-F) & 0 & 1 & 1 & 1\\\hline
$ \nu p\rightarrow lKp$ & 2&-F & -(D-F)/2 &  (D+3F) &2 &-1 & 1\\\hline
$ \nu n\rightarrow lKp$ & 1&-D-F& (D-F)/2 &  (D+3F) &1 & -2 & 0\\\hline\hline
\end{tabular}
\end{center}
\end{table*}
One can notice the induced
pseudoscalar  form factor in the $j^{\mu}\big|_{Cr\Sigma,Cr\Lambda}$ currents, which takes into account the coupling of the $W$ boson to the baryon through a kaon. However, as for the KP term, its contribution is suppressed by a factor proportional to the final lepton mass and is negligible. 
Now, we discuss in some detail the terms that appear in the coupling of the weak currents to the octet baryons in the u-channel diagrams. With very general symmetry arguments, this coupling can be described in terms of three vector and three axial form factors. Following the notation of Ref.~\cite{Cabibbo:2003cu} we have
\begin{eqnarray}
 O_V^\mu &=& f_1 \gamma^\mu+\frac{f_2}{M_B} \sigma^{\mu\nu}q_\nu+\frac{f_3}{M_B}q^\mu,\\
 O_A^\mu &=& (g_1 \gamma^\mu+\frac{g_2}{M_B} \sigma^{\mu\nu}q_\nu+\frac{g_3}{M_B}q^\mu)\gamma^5\,,
\end{eqnarray}
where $M_B$ is the baryon mass. At the order considered, the chiral Lagrangian provides finite values for $f_1$,
the weak magnetism form factor $f_2$,
$g_1$ and a pole contribution to $g_3$. The scalar $f_3$ and a non-pole part of the pseudoscalar $g_3$ form factors would only appear at higher orders of the chiral expansion. Furthermore, their contribution to the amplitude is
suppressed by a $m_l$ (lepton mass) factor and they are usually neglected.
The value of $g_2$ vanishes in the limit of  exact SU(3) symmetry and there is very little experimental information about it. In fact, it is also neglected  in most analyses of hyperon phenomenology~\cite{FloresMendieta:2004sk}.
The values of $f_1$ and $g_1$ obtained from the lowest order chiral Lagrangians describe well the hyperon semileptonic decays~\cite{Cabibbo:2003cu,FloresMendieta:2004sk,Geng:2009ik}. 

Eventually, if the cross sections for the discussed processes were measured with some precision, one could use them to explore these form factors at several $q^2$ values. The current experimental information, based on the semileptonic decays, covers only a very reduced range for this magnitude.

Finally, we consider  the $q^2$ dependence of the weak current couplings provided by the chiral Lagrangians discussed earlier. We should remark that, even at relatively low energies and low momenta of the hadrons involved in our study,  $q^2$ reaches moderate values. The $q^2$ dependences of the   needed form factors (e.g. K$\pi$, YN) are poorly known if at all. Several prescriptions have been used in the literature. For instance, for quasielastic  scattering and single pion production, the vector form factors are usually related to the well known nucleon electromagnetic ones (see e.g. \cite{Benhar:2005dj,Nieves,Leitner:2008wx} and references therein).  This procedure is well suited for these two cases because of isospin symmetry. However, in the SU(3) sector we expect to have some symmetry breaking effects. Similarly, for the axial form factors, a $q^2$ dependence obtained from the nucleon-nucleon transition obtained in neutrino nucleon quasielastic scattering is normally used. However,  the axial mass is not well established and it runs from values around 1 GeV~\cite{Bernard:2001rs,Kuzmin:2007kr}
to 1.2 GeV recently obtained by the K2K~\cite{Gran:2006jn} and MiniBooNE~\cite{:2007ru} collaborations.
Again here, we expect a different behavior for the hyperon-nucleon vertices. One of the possible choices (e.g.~\cite{Amer:1977fy}) is to use a dipole form with the mass of the vector(axial) meson that could couple the baryon to the current. In this work, in view of the present uncertainties, we adopt a global dipole form factor
$
F(q^2)=1/(1-q^2/M_F^2)^2,
$ 
with a mass $M_F\simeq 1$ GeV that multiplies the hadronic currents.  Its effect, that should be small at low neutrino energies will give an idea of the uncertainties of the calculation and will be explored in the next section.

\section{Results and Discussion}
\label{sec:res}
We consider the following reactions:
\begin{eqnarray}\label{processes}
\nu_l + p &\rightarrow & l^- + K^+ + p \;\;\;\; (l=e,~\mu)\\ \nonumber
\nu_l + n &\rightarrow & l^- + K^0 + p \\ \nonumber
\nu_l + n &\rightarrow & l^- + K^+ + n 
\end{eqnarray}
%*****************************************************************
\begin{figure}
\begin{center}
\includegraphics[width=0.82\textwidth]{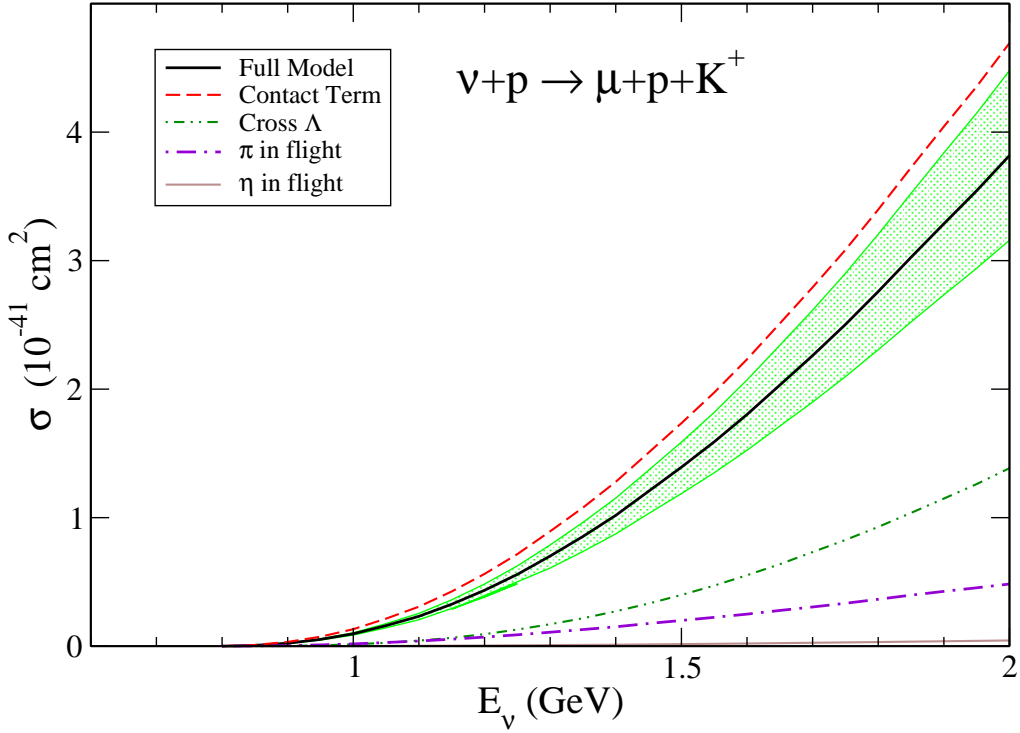}
\caption{Contribution of the different terms to the total cross section for the $ \nu_\mu p \rightarrow \mu K^+ p $ reaction.}
\label{fg:full_pp}
\end{center}
\end{figure}
%*****************************************************************
The total scattering cross section $\sigma$ has been obtained by using Eq.~(\ref{d9_sigma}) after integrating over the kinematical variables. In Figs.~(\ref{fg:full_pp}-\ref{fg:full_nn}), 
we present the results of the  contributions of the different diagrams  to the total cross sections. The kaon pole contributions are negligible at the studied energies and are not shown in the figures although they are included in the full model curves. We observe the relevance of the contact term, not included in previous calculations. Starting from the $\nu_\mu + p \rightarrow  \mu^- + K^+ + p$ channel, we find that the contact term is in fact dominant, followed by the u-channel diagram with a $\Lambda$ intermediate state and the $\pi$ exchange term. As observed by Dewan~\cite{Dewan} the u-channel $\Sigma$ contribution is much less important, basically because of the larger coupling  ($N K\Lambda \gg N K \Sigma$) of the strong vertex.
The curve labeled as Full Model has been calculated with a dipole form factor with a mass of 1 GeV. The band corresponds to changing up and down this mass by a 10 percent. A similar effect is found in the other channels and we will only show the results for the central value of 1 GeV. 
We have also checked that the cross section obtained without the contact term and after correcting for the different values of the Cabibbo angle and the Yukawa strong coupling agrees well with the result of Fig. 7 of Ref.~\cite{Dewan} at its lowest energy. Higher energies are well beyond the scope of our model.

%*****************************************************************
\begin{figure}
\begin{center}
\includegraphics[width=0.82\textwidth]{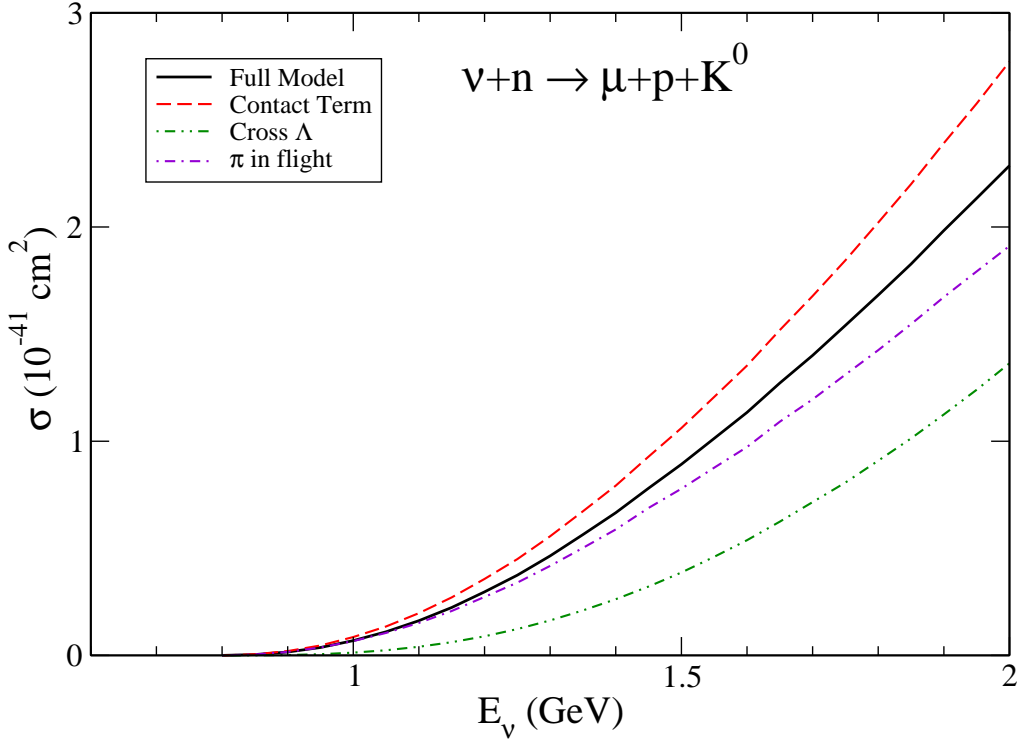}
\caption{Contribution of the different terms to the total cross section for the  $ \nu_\mu n \rightarrow \mu K^0 p $ reaction.}
\label{fg:full_np}
\end{center}
\end{figure}
%*****************************************************************
The process $\nu_\mu + n \rightarrow  \mu^- + K^0 + p$ has a cross section of a similar size and the contact term is also the largest one, followed by the $\pi$ exchange diagram and  the u-channel ($\Lambda$) term. The rate of growth of the latter is somehow larger and could become more important at higher energies.
As for the previous channel, we observe a destructive interference between the different terms and the cross section obtained with the full model is smaller than that produced by the contact term alone.

Finally, the reaction $\nu_l + n \rightarrow  l^- + K^+ + n $ has a  smaller cross section.
The pion exchange term is substantially bigger than the u-channel mechanisms, as already noted in Ref.~\cite{Dewan}. The contact term is also dominant for this channel and the total cross section
%*****************************************************************
\begin{figure}
\begin{center}
\includegraphics[width=0.82\textwidth]{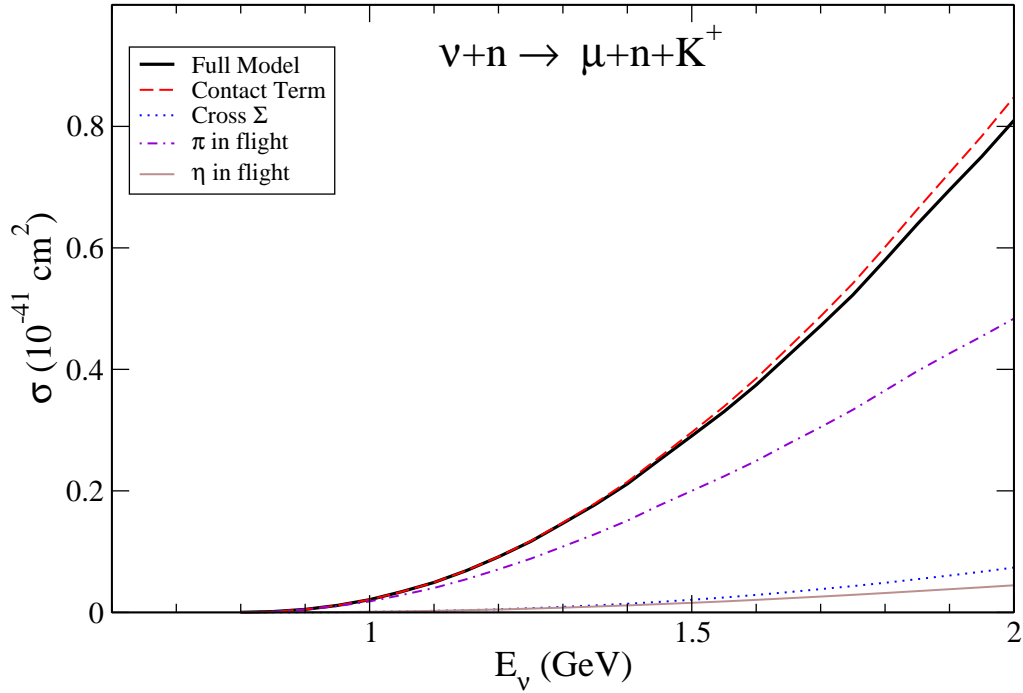}
\caption{Contribution of the different terms to the total cross section for the  $ \nu_\mu n \rightarrow \mu K^+ n $ reaction.}
\label{fg:full_nn}
\end{center}
\end{figure}
%*****************************************************************
calculated only with this term practically coincides with the full result.
Therefore, we have found that the contact terms, required by symmetry, play a major role in the description of the kaon production induced by neutrinos at low energies.

\bigskip
\begin{figure}
\begin{center}
\includegraphics[width=0.82\textwidth]{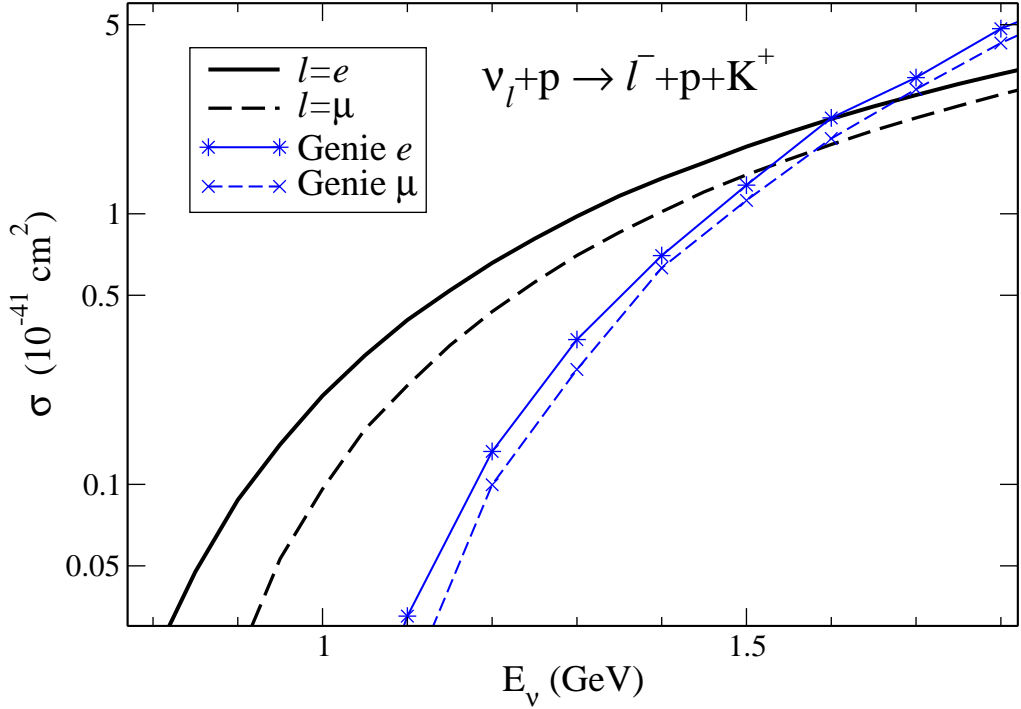}
\caption{Cross sections as a function of the neutrino energy for single kaon production vs. associated production obtained with Genie~\cite{Andreopoulos:2009rq}.}
\label{fg:genie}
\end{center}
\end{figure}
Above the energy threshold for the production of kaons accompanied by hyperons, this latter kind of processes could have larger cross sections due to the larger coupling for $\Delta S=0$, ($V_{ud}$ vs $V_{us}$). To explore this question and the range of energies where the processes we have studied are relevant we compare our results 
in Fig.~\ref{fg:genie}, with the values  for the associated production obtained by means of the GENIE Monte Carlo program~\cite{Andreopoulos:2009rq}.
We observe that, due to the difference between the energy thresholds, single kaon production for the
$\nu_l + p \rightarrow  l^- + K^+ + p$ is  clearly dominant for neutrinos of energies below 1.5 GeV. For the other two channels associated production becomes comparable at lower energies. Still,
single $K^0$ production off neutrons is larger than the associated production up to 1.3 GeV and even the much smaller $K^+$ production off neutrons is larger than the associated production up to 1.1 GeV.
The consideration of these $\Delta S=1$ channels is therefore important for the description of strangeness production for all low energy neutrino spectra and should be incorporated in the experimental analysis.

\begin{table*}
\begin{center}
\caption{Cross sections averaged over the neutrino flux at different laboratories in units of $10^{-41}$ cm$^2$. Theoretical uncertainties correspond to a 10\% variation of the form factor mass.}
\label{tab:2}
%\resizebox{148mm}{40mm}{
\begin{tabular}{|c|c c c | }\hline\hline
Process & ANL &MiniBooNE & T2K \\ 
\hline\hline
$ \nu_\mu n\rightarrow \mu^-K^+n$ & 0.06(1) &0.07(1)& 0.09(1) \\
\hline
$ \nu_\mu p\rightarrow \mu^-K^+p$ & 0.28(5) &0.32(5) & 0.43(8) \\
\hline
$ \nu_\mu n\rightarrow \mu^-K^0p$ & 0.17(3) &0.20(3)& 0.25(5) \\
\hline\hline
\end{tabular}
\end{center}
\end{table*}

In Table~\ref{tab:2} we show the total cross section results for the three channels averaged over the 
ANL~\cite{Barish:1977qk}, the MiniBooNE~\cite{AguilarArevalo:2010zc} and the off-axis (2.5 degrees) T2K~\cite{Ichikawa:2009zz} muon neutrino fluxes, all of them peaking at around 0.6 GeV. After normalization of the neutrino flux $\phi$ we have
\begin{equation}
\bar \sigma =\int_{E_{\rm th}}^{E_{\rm high}} dE\, \phi(E)
\sigma(E), 
\end{equation}
where $ E_{\rm th}$ is the threshold energy for each process and $E_{\rm high}$ is the maximum neutrino energy.
As discussed previously, in these three cases, the neutrino energies are low enough for single kaon production to be relevant as compared to associated kaon production. Also the invariant mass of the hadronic system and the transferred momentum only reach the relatively small  values where our model is more reliable. 

We can get an idea of the magnitude of these channels by comparing their cross section to some recent results. For instance, the cross section for neutral current $\pi^0$ production per nucleon has been measured by the MiniBooNE collaboration~\cite{AguilarArevalo:2009ww} obtaining $\bar{\sigma}=(4.76\pm0.05 \pm0.76 )\times 10^{-40} $ cm$^2$ with a data set of some twenty thousand valid events. The cross sections predicted by our model with the same neutrino flux are around two orders of magnitude smaller, what means that a few hundreds of kaons should have been produced. 

The atmospheric spectrum~\cite{Honda:2006qj} also peaks at very low energies and our model should be very well suited to analyse the kaon production. In Table~\ref{tab:3}, we show the number of kaon events that we obtain
%%%%%%%%%%%%%%%%%
\begin{table*}
\label{tab:3}
\begin{center}
\caption{Number of events calculated for single kaon production in water 
corresponding to the SuperK analysis for atmospheric neutrinos.}
\begin{tabular}{|c|c c  | }\hline\hline
Process &Events $e^{-}$& Events $\mu^-$\\
\hline
$\nu_l n \rightarrow l^-nK^+$ &0.16 &0.27\\
$\nu_l n \rightarrow l^-pK^0$ &0.45 &0.73\\
$\nu_l p \rightarrow l^-pK^+$ &0.95 &1.55\\
\hline
Total &1.56& 2.55 \\
\hline\hline
\end{tabular}
\end{center}
\end{table*}
%%%%%%%%%%%%%
for the 22.5 kTons of a water target and a period of 1489 days as in the SuperK analysis~\cite{Ashie:2005ik,Kobayashi:2005pe} of proton decay. As in the quoted paper, we include cuts in the electron momentum ($p_e>100$ MeV) and muon momentum ($p_e>200$ MeV). We find that single kaon production is a very small source of background. In the SuperK analysis the kaon production was modeled following Ref.~\cite{Rein:1980wg,Rein:1987cb} and only included associated kaon production. Although some of the cuts applied in their analysis, such as looking for an accompanying hyperon, are useless for our case,
we find that  this source of background is negligible,  given the smallness of our results and the totally different energy distribution of kaons and final leptons in the production and decay reactions. 

\begin{figure}
\begin{center}
\includegraphics[width=\textwidth]{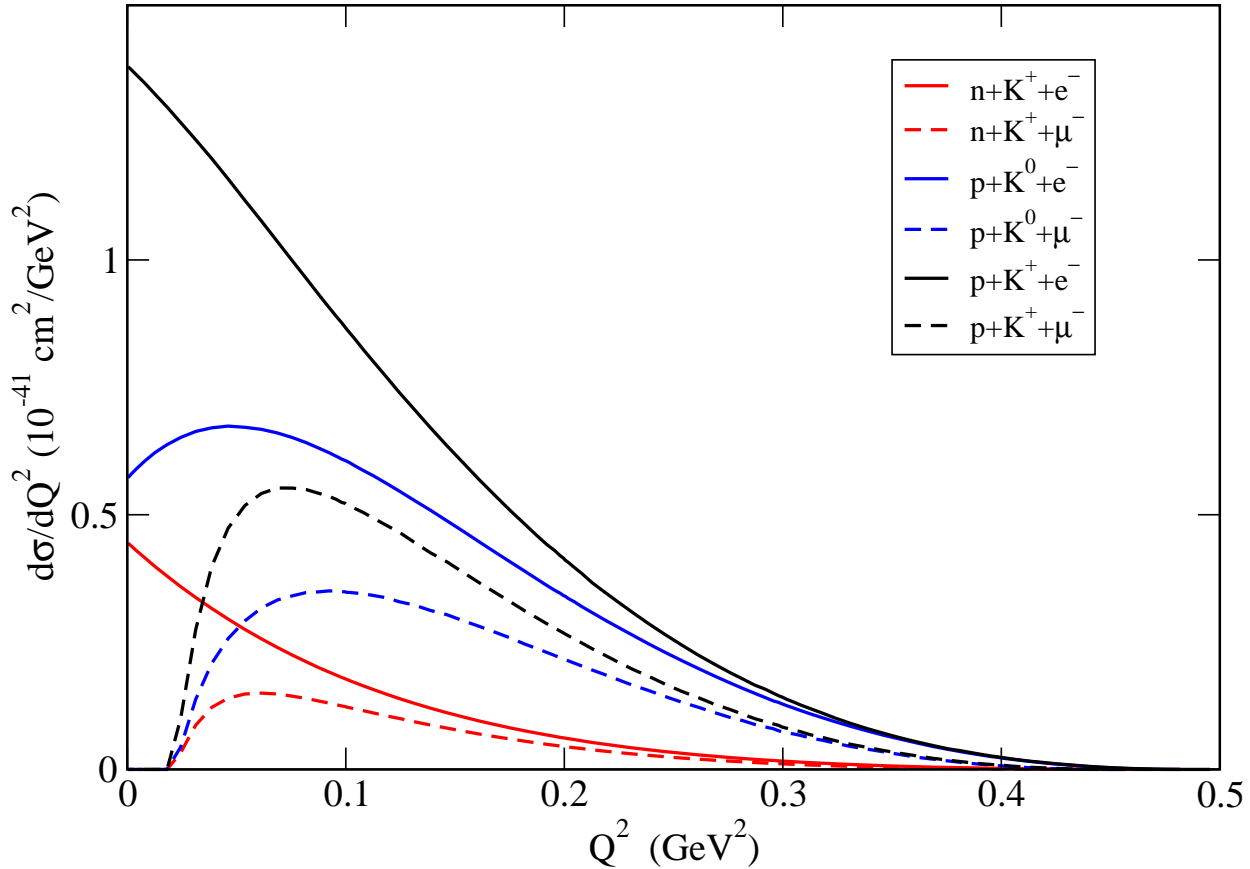}
\caption{$\frac{d \sigma}{ d Q^2} $ at $ E_\nu=1 GeV $ for single kaon production induced by neutrinos. The curves are labeled according to the final state of the process.}
\label{fg:dsdq2}
\end{center}
\end{figure}

Finally, we study the values of $Q^2$ involved in the reaction for the typical neutrino energies we have considered. If high values of this magnitude are relevant, the results would be sensitive to higher orders of the chiral Lagrangians and/or a more precise description of the form factors. We show the 
$Q^2$ distribution   in Fig.~(\ref{fg:dsdq2}) for the three studied channels at a neutrino energy E$_\nu= 1$ GeV. The reactions are always forward peaked (for the final lepton), even in the absence of any form factor ($F(q^2)=1$), favouring relatively small values of the momentum transfer.
%*****************************************************************
%*****************************************************************
In this figure, we also show the dependence of the cross section on the mass of the final lepton that reduces the  cross section at low $Q^2$ values. The process  $\nu_e + n \rightarrow  e^- + K^0 + p$ shows a slightly different behavior that reflects an important (and $Q^2$ dependent) interference between the pion exchange and the contact terms. 

Till now we have discussed the kaon production off free nucleons. However, most of the experiments are carried out 
on detectors containing complex nuclei such as iron, oxygen or carbon. On the other hand, nuclear effects are known to be quite large for pion production induced by neutrinos~\cite{AlvarezRuso:1998hi,Leitner:2006ww,Leitner:2008wx,Amaro:2008hd}.
Fortunately, this question is much simpler for the kaons. First, because there is no kaon absorption and the final state interaction is reduced to a repulsive potential, small when compared with the typical kaon energies. Second, because of the absence of resonant channels in the production processes. We could remember here that some of the major nuclear effects for pion production are originated by the modification of the $\Delta(1232)$ properties on nuclei. Other nuclear effects, such as Fermi motion and Pauli blocking will only produce minor changes on the cross  section and can easily be implemented in the Monte Carlo codes.

In summary, we have developed a microscopical model for  single kaon production off nucleons induced by neutrinos based on the SU(3) chiral Lagrangians. This model should be quite reliable at low and intermediate energies given the absence of  $S=1$ baryonic resonances in the s-channel. The parameters of the model are well known:
$f_\pi$, the pion decay constant,  Cabibbo's angle, the proton and neutron magnetic moments and the  axial vector coupling constants for the baryons octet, $D$ and $F$. For the latter ones, we have taken the values obtained from the analysis of the hyperon semileptonic decays. The importance of higher order terms has been estimated using a dipole form factor with a mass around 1GeV and exploring the dependence of our results on this parameter.

We obtain cross sections that are around two orders of magnitude smaller than for pion production for neutrino spectra such as those of ANL or MiniBooNE. This can be understood because of the Cabibbo suppression and of the smaller phase space. Nonetheless, the cross sections are large enough to be measured, for instance,  with the expected Minerva and T2K fluxes and could have been well measured at MiniBooNE.
We have also found, that due to the higher threshold of the associated kaon production, the reactions we have studied are the dominant source of kaons for a wide range of energies, and thus their study is important for some low energy experiments and for the atmospheric neutrino flux.

\begin{acknowledgments}
We want to acknowledge discussions with P. Stamoulis.
This work is partly supported by DGICYT Contract No. FIS
2006-03438, the Generalitat Valenciana in the program Prometeo
and the EU Integrated Infrastructure Initiative Hadron
Physics Project under contract RII3-CT-2004-506078. I.R.S. acknowledges support from
the Ministerio de Educaci\'on.M.R.A. wishes to acknowledge the financial support from the University of Valencia and Aligarh Muslim
University under the academic exchange program and also to the DST, Government of India for the financial support
under the grant SR/S2/HEP-0001/2008.
\end{acknowledgments}

\end{document}